\documentclass[prd,preprint,nofootinbib]{revtex4}
\usepackage{amssymb}
\usepackage{epsfig}

\begin{document}

\title{ 
 A brane world cosmological solution to the gravitino problem 
}

\author{Nobuchika Okada}
 \email{okadan@post.kek.jp}
 \affiliation{
  Theory Division, KEK, Oho 1-1, Tsukuba, Ibaraki 305-0801, Japan
 }

\author{Osamu Seto}
 \email{osamu@mail.nctu.edu.tw}
 \affiliation{
 Institute of Physics, National Chiao Tung University, 
 Hsinchu, Taiwan 300, Republic of China 
}

\begin{abstract}
We investigate the thermal production of gravitinos 
 in the context of the brane world cosmology. 
Since the expansion law is modified 
 from the one in the standard cosmology, 
 the Boltzmann equation for the gravitino production is altered. 
We find that the late-time gravitino abundance 
 is proportional to the ``transition temperature'', 
 at which the modified expansion law in the brane world cosmology 
 is connecting with the standard one, 
 rather than the reheating temperature after inflation 
 as in the standard cosmology. 
This means that, 
 even though the reheating temperature is very high, 
 we can avoid the overproduction of gravitinos 
 by taking the transition temperature low enough. 
Therefore, the gravitino problem can be solved. 
\end{abstract}


\vspace*{3cm}
\maketitle


Several particles which only gravitationally couple 
 to the ordinary matters appear in supergravity models 
 (some models inspired by string theories). 
Since their couplings are Planck suppressed, 
 their lifetime is quite long as estimated to be 
\begin{equation}
 \tau \simeq \frac{M_{Pl}^2}{m^3} 
 \simeq 10^5 \left(\frac{1\textrm{TeV}}{m}\right)^3 
 \textrm{sec},  
\end{equation}
where $m$ is the mass of the particles, and 
 $M_{Pl} \simeq 1.2 \times 10^{19}$ GeV is the Planck mass. 
It is normally assumed that 
 the scale of $m$ is of order of the electroweak scale 
 which is obtained through supersymmetry breaking 
 in supergravity models. 
These particles with such a long lifetime often cause 
 a serious problem in cosmology. 
The so-called ``gravitino problem'' is a famous example 
 \cite{Khlopov}. 
If the gravitino decays after big bang nucleosynthesis (BBN), 
 its energetic daughters would destroy the light nuclei 
 by photo-dissociation and, as a result, 
 upset the successful prediction of BBN. 
As is well known, to avoid this problem 
 the upper bound (depending on the gravitino mass) is imposed 
 on the reheating temperature after inflation. 

Here let us see a little more detail of the gravitino problem. 
Gravitinos have been produced by thermal scatterings in plasma
 since the reheating stage after inflation. 
The Boltzmann equation relevant to this gravitino production process
 is given by
\begin{equation}
\frac{d n_{3/2}}{d t}+3H n_{3/2} = 
 \langle\sigma_{tot} v\rangle n_{rad}^2,
\label{Boltz:n}
\end{equation}
where $H$ is the Hubble parameter, $n_{3/2}$ and $n_{rad}$ are
 the number densities of gravitino and relativistic species  
 in thermal bath, respectively, 
 and $\langle \sigma_{tot} v \rangle$ denotes 
 the thermal average of the total gravitino production cross section 
 times the relative velocity. 
In the standard cosmology, 
 the Hubble parameter obeys the Friedmann equation,  
\begin{eqnarray}
 H^2 = \frac{8\pi G}{3}\rho, 
 \label{FriedmannEq} 
\end{eqnarray}
with the energy density $\rho = \frac{\pi^2}{30}g_*T^4 $ 
 in the radiation dominated universe. 
Here, $G$ is the Newton's gravitational constant, 
 and $g_*(T)$ is the total number of the massless 
 degrees of freedom.
It is useful to rewrite Eq.~(\ref{Boltz:n}) as 
\begin{equation}
\frac{d Y_{3/2}}{d T} = 
 - \frac{s\langle\sigma_{tot} v\rangle}{HT}Y_{rad}^2,
\label{Boltz:Y}
\end{equation}
where $Y_{3/2}\equiv n_{3/2}/s$ and $Y_{rad}\equiv n_{rad}/s$ 
 with $s$ being the entropy density. 
Note that the r.h.s. of this equation is almost independent of 
 temperature $T$, 
 since $ \langle \sigma_{tot} v \rangle \simeq M_{Pl}^{-2}$, 
 $H \propto T^2$, $s \propto T^3$, etc. 
Thus we find that the late-time gravitino abundance 
 is proportional to the reheating temperature $T_R$ 
 and approximately described as \cite{KawasakiMoroi},
\begin{eqnarray}
 Y_{3/2} \simeq 10^{-11} 
 \left(\frac{T_R}{10^{10} \textrm{GeV}}\right). 
 \label{YresultSTD}
\end{eqnarray}
This abundance should be small 
 in order for the gravitino decay products 
 not to destroy the light elements successfully synthesized 
 during BBN, 
 thus we obtain the upper bound on the reheating temperature 
 such as \cite{Cyburt,KKM} 
\begin{eqnarray}
 T_R \leq  10^6-10^7 \textrm{GeV}, 
 \label{bound}
\end{eqnarray}
for the gravitino mass 
 $m_{3/2} = \mathcal{O}(10^2 \mbox{GeV})$, 
 for example. 
In the recent analysis \cite{KKM} 
 the effect of the hadronic processes also has been taken into account. 

This stringent constraint is problematic in the inflationary cosmology. 
The reheating temperature provided by typical inflation models  
 such as the chaotic inflation or the hybrid inflation 
 is estimated to be much larger 
 than the upper bound of Eq.~(\ref{bound}) \cite{NewRef}. 
Therefore there exists a gap 
 between the natural reheating temperature 
 for the inflation models 
 and its upper bound of Eq.~(\ref{bound}). 
It seems not so easy to construct an inflation model 
 which can, simultaneously, provide a reheating temperature 
 low enough to satisfy this upper bound  
 and be consistent with the present cosmological observations, 
 the density fluctuations, etc. 
\footnote{ 
Some inflation models with low reheating temperature 
 has been proposed in the context of the brane world cosmology 
 \cite{Mazumdar, Bento}, 
 where some issues related to this letter are also discussed. } 
Some ideas to avoid the gravitino problem have been proposed 
 such as the thermal inflation \cite{ThermalInf} 
 and a model to forbid the radiative decay of gravitinos \cite{AY}. 

Recently, the brane world models have been attracting
 a lot of attention as a novel higher dimensional theory. 
In these models, it is assumed that the standard model particles 
 are confined on a ``3-brane'' 
 while gravity resides in the whole higher dimensional spacetime.
The model first proposed by Randall and Sundrum \cite{RS}, 
 the so-called ``RS II'' model, is a simple and interesting one, 
 and its cosmological evolutions have been 
 intensively investigated \cite{braneworld}. 
In the model, our 4-dimensional universe is realized 
 on the 3-brane with a positive tension 
 located at the ultraviolet (UV) boundary 
 of a five dimensional Anti de-Sitter spacetime. 
In this setup, the Friedmann equation for a spatially flat spacetime 
 is found to be 
\begin{equation}
H^2 = \frac{8\pi G}{3}\rho\left(1+\frac{\rho}{\rho_0}\right) 
\label{BraneFriedmannEq}
\end{equation}
where
\begin{eqnarray}
\rho_0 = 96 \pi G M_5^6, 
\end{eqnarray}
with $M_5$ being the five dimensional Planck mass, 
 and the four dimensional cosmological constant has been tuned 
 to be zero. 
Here we have omitted the term so-called ``dark radiation'', 
 since this term is severely constrained by big bang nucleosynthesis 
 \cite{ichiki}.
The second term proportional to $\rho^2$ 
 is a new ingredient in the brane world cosmology 
 and lead to a non-standard expansion law. 

Note that according to Eq.~(\ref{BraneFriedmannEq}) 
 the evolution of the early universe can be divided into two eras. 
At the era where $\rho/\rho_0 \gg 1$ the second term dominates 
 and the expansion law is non-standard (brane world cosmology era), 
 while at the era $\rho/\rho_0 \ll 1$ the first term dominates 
 and the expansion of the universe 
 obeys the standard expansion law (standard cosmology era). 
In the following, we call a temperature defined as 
 $\rho(T_t)/\rho_0=1$ ``transition temperature'', 
 at which the evolution of the early universe 
 changes from the brane world cosmology era 
 into the standard one.
\footnote{
We can find a lower bound on the transition temperature 
 such as $T_t > {\cal{O}} (1  \mbox{TeV})$ 
 according to the discussion in \cite{RS} 
 concerning the precision measurements of 
 the Newton's gravitational law in sub-millimeter range. }
Using the transition temperature, 
 we rewrite Eq.~(\ref{BraneFriedmannEq}) into the form 
\begin{eqnarray}
H^2 = \frac{8\pi G}{3}\rho 
 \left[ 1+ \left(\frac{T}{T_t} \right)^4 \right]  
 = H_{st}^2 \left[1+ \left(\frac{T}{T_t} \right)^4 \right] , 
\label{BraneFriedmannEqII}
\end{eqnarray}
where $H_{st}$ is the Hubble parameter in the standard cosmology 
 (Eq.~(\ref{FriedmannEq})). 

This modification of the expansion law at a high temperature 
 ($T > T_t$) leads some drastic changes 
 for several cosmological issues. 
In fact, for example, an enhancement of the thermal relic abundance 
 of dark matter has been recently pointed out \cite{OS}. 
In this letter, we investigate the brane cosmological effect 
 for the gravitino production process. 
Since the Friedmann equation is modified into 
 Eq.~(\ref{BraneFriedmannEqII}), 
 the solution of the Boltzmann equation of Eq.~(\ref{Boltz:Y}) 
 can be modified, 
 which leads us to a different consequence 
 for the gravitino problem. 

Before proceeding our discussion, let us specify our setup. 
Our brane world model is 
 the supersymmetric extended RS II model \cite{RSsugra}. 
The cosmological solution of this extended model is the same 
 as that in the non-supersymmetric model, 
 since the Einstein's equation belongs to the bosonic part. 
As is well known in the RS II model, 
 the zero-mode graviton is localized around the UV brane, 
 while the graviton Kaluza-Klein (KK) modes are likely localized 
 around the infrared (IR) brane 
 (in the RS II model, the IR brane is located at the infinite boundary). 
This feature is the key to recover 
 almost four-dimensional Einstein gravity 
 even with the infinite fifth dimensional radius, 
 namely, the idea of the ``alternative to compactification''. 
If the supersymmetry is manifest, all the component fields 
 in a supergravity multiplet take the same configurations 
 in the fifth dimensional direction. 
Thus the zero-mode gravitino is localized around the UV brane, 
 and as an approximation we regard it 
 as a field residing on the UV brane. 
Suppose that the SUSY is broken on the UV brane. 
The zero-mode gravitino obtains its mass through 
 the super-Higgs effect as in four dimensional supergravity,
 while KK gravitinos would remain the same, 
 since supersymmetry is broken only locally 
 on the UV brane in the infinite fifth dimensional coordinate. 
As discussed above, we treat the zero-mode gravitino as the field 
 on the UV brane (the brane on which 
 all the supersymmetric standard model fields reside) 
 and apply the above Boltzmann equation 
 for the gravitino production.  
On the other hand, the plasma on the UV brane (our brane) 
 can, in general, excite the KK gravitinos 
\footnote{The KK graviton can also be emitted from the plasma 
 on the UV brane and contribute to the dark radiation term 
 in the Friedmann equation. 
 This issue has been studied in \cite{KK-graviton}, 
 and the resultant dark radiation is found to be consistent 
 with the BBN constraint.}. 
Since the KK gravitinos take different configurations 
 in the fifth dimensional direction 
 and couple differently to the fields on our brane, 
 their production processes will obey 
 different types of Boltzmann equations. 
It is very interesting and challenging 
 to analyze the KK gravitino production process. 
We leave this important issue for future works. 
 
Now let us analyze the thermal production of the gravitino 
 in the brane world cosmology. 
The Boltzmann equation we analyze takes the form of 
\begin{eqnarray}
\frac{d Y_{3/2}}{d T} = 
 - \frac{s\langle\sigma_{tot} v\rangle}{H_{st}T}
 Y_{rad}^2 \times 
 \frac{1}{\sqrt{1+\left(\frac{T}{T_t}\right)^4}}. 
\label{BraneBoltz:Y}
\end{eqnarray}
The pre-factor in the r.h.s. is the same 
 as the one in the standard cosmology. 
Suppose that the reheating temperature after inflation 
 is much higher than the transition temperature. 
Then we integrate the Boltzmann equation 
 from the reheating temperature $T_R (\gg T_t)$ 
 in the brane world cosmology era 
 to a low temperature $T_{low} (\ll T_t)$
 in the standard cosmology era. 
First let us integrate the Boltzmann equation 
 approximately in such a way that 
\begin{eqnarray}
- \int_{T_R}^{T_{low}}  
 \frac{d T}{\sqrt{1+\left(\frac{T}{T_t}\right)^4}} 
 \sim  
 \int^{T_R}_{T_t} dT \left(\frac{T_t}{T} \right)^2 
 + \int^{T_t}_{T_{low}}  dT  \sim  2 T_t.    
\label{BraneBoltz:Yapp}
\end{eqnarray}
We will see this result is a good approximation 
 from the following analytic formula we will show. 
Note that the reheating temperature $T_R$  
 in the formula of the gravitino abundance 
 in the standard cosmology (see Eq.~(\ref{YresultSTD})) 
 is replaced by the transition temperature, 
 namely, $T_R \rightarrow 2 T_t$. 
Therefore we can avoid overproduction of gravitinos by 
\begin{eqnarray}
 T_t  \leq  10^6- 10^7 \mbox{GeV} 
 \label{boundBrane}
\end{eqnarray}
 independently of the reheating temperature. 
This is our main result in this letter. 
Needless to say, if $T_t \gg T_R$ we can reproduce the result 
 in the standard cosmology. 

For completeness, we show analytic formulas 
 of the solution of the Boltzmann equation as follows: 
\begin{eqnarray}
 Y_{3/2}(T_{low})
 & = & \int_{T_{low}}^{T_R} 
 \frac{s\langle\sigma_{tot} v\rangle}{H_{st}T}Y_{rad}^2
 \frac{dT}{\sqrt{1+\left(\frac{T}{T_t}\right)^4}}
 \nonumber \\ 
&\simeq& 
 \left.\frac{s\langle\sigma_{tot} v\rangle}{H_{st}T}Y_{rad}^2\right|_{T=T_R} T_t 
 \left[ 
  Z_R F \left( 
   \frac{1}{4}, \frac{1}{2}, \frac{5}{4}; -Z_R^4  
  \right)  \right]  
\end{eqnarray}
where $F(\alpha,\beta,\gamma;z)= {}_2F_1(\alpha,\beta,\gamma;z)$ 
 is the Gauss' hypergeometric function, 
 and $Z_R$ is defined as $Z_R=T_R/T_t (\gg 1)$. 
By using the relations, 
\begin{eqnarray}
 F(\alpha,\beta,\gamma;z) &=&
 (1-z)^{-\alpha} F
 \left(
  \alpha,\gamma-\beta,\gamma; \frac{z}{z-1}  
 \right) ,  
 \nonumber \\
 F(\alpha,\beta,\gamma;1) &=& 
 \frac{\Gamma(\gamma)\Gamma(\gamma-\alpha-\beta)}
 {\Gamma(\gamma-\alpha)\Gamma(\gamma-\beta)},  
\end{eqnarray}
we obtain
\begin{equation}
 Z_R F \left(\frac{1}{4}, \frac{1}{2}, \frac{5}{4}; -Z_R^4 \right)
 \longrightarrow 
 \frac{\Gamma(5/4)\Gamma(1/4)}{\Gamma(1)\Gamma(1/2)} 
 \simeq 1.85  
 \end{equation}
for $Z_R \gg 1$. 
This result shows that Eq.~(\ref{BraneBoltz:Yapp}) 
 is actually a good approximation. 

If the reheating is completed in the brane world cosmology era 
 as we have assumed, 
 the reheating temperature becomes different 
 from the one in the standard cosmology 
 due to the non-standard expansion law. 
Here let us estimate the reheating temperature 
 and check that our assumption, $T_R \gg T_t$, can be satisfied. 
Comparing the total decay width of an inflaton ($\Gamma_I$)
 and the Hubble parameter such as 
\begin{eqnarray} 
 \Gamma_I^2 = H^2 = \frac{8 \pi G}{3} \rho(T_R)  
  \left[1+ \left(\frac{T_R}{T_t}\right)^4  \right] 
\simeq 
 \frac{T_R^8 }{M_P^2 T_t^4} ,  
\end{eqnarray}
where $M_P=M_{Pl}/\sqrt{8 \pi}$ is the reduced Planck mass, 
 we find 
\begin{eqnarray} 
T_R \simeq T_t 
  \left( \frac{M_P}{T_t} \right)^{1/4} 
  \left( \frac{\Gamma_I}{T_t} \right)^{1/4}  .
\label{ReheatBrane}
\end{eqnarray}
Therefore our assumption can be satisfied 
 if $(M_P \Gamma_I/T_t^2)^{1/4} \gg 1$. 
It is interesting to see a relation 
 between the reheating temperature in the brane world cosmology 
 and the standard one. 
In the standard cosmology, the reheating temperature is estimated 
 as $T_R^{(s)} \simeq \sqrt{M_P \Gamma_I}$. 
Substituting this into Eq.~(\ref{ReheatBrane}), 
 the reheating temperature of the brane world cosmology 
 is described as 
\begin{eqnarray}    
T_R  \simeq T_t \left( \frac{T_R^{(s)}}{T_t}
\right)^{1/2} . 
\end{eqnarray} 
Recall that the gap 
 between $T_R^{(s)}$ and the upper bound of Eq.~(\ref{bound}),
 namely, $T_R^{(s)} \gg 10^6-10^7 \mbox{GeV}$ 
 causes the gravitino problem originally. 
As discussed above, Eq.~(\ref{bound}) is replaced 
 by Eq.~(\ref{boundBrane}) in the brane world cosmology. 
Therefore the gap between the temperatures, 
 on the contrary, justifies 
 our assumption $T_R \gg T_t$ due to $T_R^{(s)} \gg T_t$. 

In summary, we investigate 
 the abundance of gravitinos 
 in the framework of inflationary brane world cosmology. 
The point is that the expansion law is modified 
 in the brane world cosmology 
 and this modification can lead to different results
 for various issues in the standard cosmology. 
We have found that the late-time gravitino abundance 
 is proportional to the transition temperature 
 independently of the reheating temperature, 
 if the reheating temperature is higher 
 than the transition temperature. 
Therefore the gravitino problem disappears 
 by assigning a transition temperature low enough. 
We also have checked that consistency of our assumption, 
 $T_t \ll T_R$. 

Finally we give some comments. 
A necessary condition to solve the gravitino problem by this manner 
 is that inflation must occur in the brane world cosmology era 
 as we have assumed. 
In order to verify our strategy 
 toward a solution of the gravitino problem, 
 it is important to explore whether the inflation can take place 
 during the brane world cosmology era \cite{InflationOnBrane}.
Furthermore, the detailed study of density perturbation 
 generated from inflation models in the brane world cosmology 
 is an important future issue \cite{Perturbation1,Perturbation2}. 
Also, as mentioned above, 
 the issue of the KK gravitino production 
 should be investigated in future.

%
N.O. would like to thank 
 the Abdus Salam International Centre 
 for Theoretical Physics (ICTP), Trieste, 
 during the completion of this work. 
The work of N.O. is supported in part 
 by the Grant-in-Aid for Scientific Research  (\#15740164). 
O.S. is supported by the National Science Council of Taiwan 
under the grant No. NSC 92-2811-M-009-018.



\end{document}